\documentstyle[preprint,aps,graphicx]{revtex}

\begin{document}
\draft
\title{{Constraints from Neutrinoless Double Beta Decay}}

\author{K. MATSUDA, T. KIKUCHI, T. FUKUYAMA}
\address{
Department of Physics, 
Ritsumeikan University, Kusatsu, Shiga, 525-8577 Japan}
\author{H. NISHIURA}
\address{
Department of General Education, 
Junior College of Osaka Institute of Technology, \\
Asahi-ku, Osaka,535-8585 Japan}

\date{April 15, 2002}
\maketitle

\begin{abstract}

We examine the constraints from 
the recent HEIDELBERG-MOSCOW double beta decay experiment.
It leads us to the almost degenerate or inverse hierarchy neutrino mass scenario. In this scenario,  we obtain possible upper bounds for the Majorana $CP$ violating phase in the lepton sector by incorporating the data from the neutrino oscillation, the single beta decay experiments, and from the astrophysical observation. 
We also predict the neutrino mass that may be measurable in the future beta decay experiments. 
\end{abstract}
The recent neutrino oscillation experiments\cite{skamioka} 
have shown that neutrinos have masses.
On the 
other hand, the experiments intending to determine directly 
neutrino mass such as the neutrinoless double beta decay($(\beta\beta)_{0\nu}$) and the single beta decay experiments are also on going. In a series of papers, we have discussed the $CP$ violation effects in the lepton sector incorporating all these experiments \cite{Matsuda} and the other indirect astrophysical observations \cite{nishiura}.
Recently, Klapdor-Kleingrothaus et al.\cite{Klapdor} argued 
the evidence for $(\beta\beta)_{0\nu}$ by analyzing the data of the HEIDELBERG-MOSCOW experiment and reported that
\begin{equation}
0.11<\langle m_{\nu} \rangle <0.56\mbox{eV (95$\%$ C.L.)}\label{HEIDELBERG-MOSCOWmass}
\end{equation}
with the best fit value, \(\langle m_{\nu} \rangle =0.39\) eV.  
\par
In this paper we reanalyze our studies in response to this announcement, although some papers\cite{Klapdor2} have discussed the constraints from this data. We especially consider the $CP$ violation effects in it. 
Namely, using the data of Eq.(\ref{HEIDELBERG-MOSCOWmass}), we try to derive the constraints on the $CP$ violating phases in the lepton sector by combining the constraints from the neutrino oscillations, the beta decay experiments, and the astrophysical observations. Since the direct test of $CP$ violations from a measurement of such as electric dipole moments of leptons seems to be infeasible at present, the $(\beta\beta)_{0\nu}$ can be a good channel to detect the $CP$ violation effects, although they are indirectly measurable.
\par
The $(\beta\beta)_{0\nu}$ experiment 
gives us the information of the averaged mass $\langle m_{\nu} \rangle$ for Majorana neutrinos defined by\cite{doi} 
\begin{equation}
\langle m_{\nu} \rangle\equiv |\sum _{j=1}^{3}U_{ej}^2m_j|
=|m_1|U_{e1}|^2+m_2|U_{e2}|^2e^{2i\beta}+m_3|U_{e3}|^2e^{2i\rho\prime}|.\label{betabetamass}
\end{equation}
Here \(\beta\) and \(\rho\prime\equiv \rho-\phi\) are $CP$ violating phases. 
The $U_{a j}$ is the Maki-Nakagawa-Sakata (MNS)\cite{MNS} left-handed lepton mixing matrix that combines the weak eigenstate neutrino 
($a=e,\mu$ and $\tau$) with the mass eigenstate neutrino of 
mass $m_j$ ($j$=1,2 and 3).  
The $U$ takes the following form in the standard representation \cite{Matsuda}:
\begin{equation}
U=
\left(
\begin{array}{ccc}
c_1c_3&s_1c_3e^{i\beta}&s_3e^{i(\rho-\phi )}\\
(-s_1c_2-c_1s_2s_3e^{i\phi})e^{-i\beta}&
c_1c_2-s_1s_2s_3e^{i\phi}&s_2c_3e^{i(\rho-\beta )}\\
(s_1s_2-c_1c_2s_3e^{i\phi})e^{-i\rho}&
(-c_1s_2-s_1c_2s_3e^{i\phi})e^{-i(\rho-\beta )}&c_2c_3\\
\end{array}
\right).\label{CKM}
\end{equation}
Here $c_j=\cos\theta_j$, $s_j=\sin\theta_j$ 
($\theta_1=\theta_{12},~\theta_2=\theta_{23},~\theta_3=\theta_{31}$). 
Note that three 
$CP$ violating phases, $\beta$ , $\rho$ and $\phi$ appear in $U$ for 
Majorana particles \cite{bilenky}. 

The other experimental constraints on neutrino mass and neutrino mixing angles are as follows:
The recent beta decay experiments \cite{weinheimer} restrict another averaged neutrino mass $\overline{m_{\nu}}$ as
\begin{equation}
\overline{m_{\nu}} \equiv  \sqrt{\sum _{j=1}^{3}|U_{ej}|^2m_j^2} <2.2\mbox{ eV}.\label{betamass}
\end{equation}
From the solar neutrino oscillation experiment\cite{skamioka}, we have
\begin{eqnarray} 
&&\Delta m_{12}^2=m_2^2-m_1^2=\Delta m_{\mbox{\tiny solar}}^2=(2-20)\times 10^{-5}\mbox{eV}^2,
\label{eq41301} \\
&&0.3 \le \sin^2{2\theta\mbox{\tiny solar}}\le0.93   \quad\mbox{for LMA-MSW}, \label{eq20501} 
\end{eqnarray} 
and
\begin{eqnarray} 
&& \Delta m_{12}^2=m_2^2-m_1^2=\Delta m_{\mbox{\tiny solar}}^2=(4-9)\times 10^{-6}\mbox{eV}^2, \\
&& 8 \times 10^{-4}\le \sin^2 {2\theta_{\mbox{\tiny solar}}} \le 8 \times 10^{-3}   
\quad\mbox{for SMA-MSW}, \label{eq20502}
\end{eqnarray}
for the large mixing angle (LMA) and small mixing angle (SMA) MSW solutions, respectively.
From the atmospheric neutrino oscillation experiment\cite{skamioka}, we obtain
\begin{equation}
\Delta m_{23}^2=m_3^2-m_2^2=\Delta m_{\mbox{{\tiny atm}}}^2=
\Biggl\{ 
\begin{array}{rl}
 (1-7)\times 10^{-3}\mbox{eV}^2  & \mbox{for normal hierarchy case }\\ 
-(1-7)\times 10^{-3}\mbox{eV}^2  & \mbox{for inverse hierarchy case}. 
\end{array}
\label{eq41302}
\end{equation}
The astrophysical observation \cite{fukugita} gives 
\begin{equation}
\sum_i m_i < 1.8 \mbox{ eV}\label{astromass}
\end{equation}
under some reasonable assumptions.
\par
Hereafter we denote the experimental lower and upper bounds in Eq.(\ref{HEIDELBERG-MOSCOWmass}) as \(\langle m_{\nu} \rangle\mbox{\tiny min}\) and \(\langle m_{\nu} \rangle\mbox{\tiny max}\), respectively. (\(\langle m_{\nu} \rangle\mbox{\tiny min}\le \langle m_{\nu} \rangle \le \langle m_{\nu} \rangle\mbox{\tiny max}\)).
Let us first show that
the data of the HEIDELBERG-MOSCOW $(\beta\beta)_{0\nu}$ experiment in Eq.(\ref{HEIDELBERG-MOSCOWmass}) prefers the almost degenerate or inverse hierarchy neutrino mass scenario for the LMA-MSW solution: 
Irrespectively of the $CP$ violating phases Eq.(\ref{betabetamass}) leads to the inequality that 
\begin{equation}
\langle m_{\nu} \rangle\le  m_1|U_{e1}|^2+m_2|U_{e2}|^2+m_3|U_{e3}|^2.
\label{eq041501}
\end{equation} 
Since we have the constraint \(|U_{e3}|^2<0.03\) from the oscillation experiments of CHOOZ\cite{chooz} and SuperKamiokande\cite{skamioka}, Eq.(\ref{eq041501}) becomes
\begin{equation}
\langle m_{\nu} \rangle< |U_{e1}|^2m_1+|U_{e2}|^2\sqrt{m_1^2+\Delta m_{12}^2}
+0.03\sqrt{m_1^2+\Delta m_{12}^2+\Delta m_{23}^2}.
\label{hierarchy}
\end{equation}
It is apparent from Eqs.(\ref{HEIDELBERG-MOSCOWmass}), (\ref{eq41301}), (\ref{eq41302}), and (\ref{hierarchy}) that the normal hierarchy, \(m_1\alt m_2\ll m_3\), is forbidden.
We have no way of distinguishing between the almost degenerate and inverse hierarchy neutrino mass scenarios based on Eq.(\ref{HEIDELBERG-MOSCOWmass}) at this stage. Hence we obtain 
\begin{eqnarray}
\langle m_{\nu} \rangle& \simeq& m||U_{e1}|^2+|U_{e2}|^2e^{2i\beta}|,\label{betabetamass2}\\
\overline{m_{\nu}}& \simeq& m, \label{betamass2}
\end{eqnarray}
with \(m\equiv m_1 \simeq m_2\).
Since \(|U_{e3}|^2<0.03\), \(\sin^2{2\theta\mbox{\tiny solar}}\) becomes \(4|U_{e2}|^2(1-|U_{e2}|^2)\) and Eq.(\ref{betabetamass2}) is rewritten as
\begin{equation}
\sin^2\beta=\frac{1}{\sin^2{2\theta\mbox{\tiny solar}}}
\left(1-\frac{\langle m_{\nu} \rangle^2}{m^2}\right) .\label{betaphase}
\end{equation}
For LMA-MSW solution, Eq.(\ref{betaphase}) gives
\begin{equation}
\frac{1}{(\sin^2{2\theta\mbox{\tiny solar}})\mbox{\tiny max}}
\left(1-\frac{\langle m_{\nu} \rangle^2}{m^2}\right) \le \sin^2\beta \le \frac{1}{(\sin^2{2\theta\mbox{\tiny solar}})\mbox{\tiny min}}
\left(1-\frac{\langle m_{\nu} \rangle^2}{m^2}\right) .\label{eq05}
\end{equation}
Here we have denoted the experimental lower and upper limits of Eq.(\ref{eq20501}) as \((\sin^2{2\theta\mbox{\tiny solar}})\mbox{\tiny min}\) 
and \((\sin^2{2\theta\mbox{\tiny solar}})\mbox{\tiny max}\), respectively.
The allowed region in the $\sin^2\beta-m$ plane for the LMA-MSW is shown by the shaded area in Fig.1 with use of the experimental bounds in Eq.(\ref{HEIDELBERG-MOSCOWmass}). 
Let us superimpose on this allowed region the constraint of 
the experimental upper bound \(m\mbox{\tiny max}\) for \(m\) that is obtained from astrophysical observation and single beta decay. We have \(m< 0.6 \mbox{ eV}\) from Eq.(\ref{astromass}) and \(m<2.2\mbox{ eV}\) from Eq.(\ref{betamass}) using \(\overline{m_{\nu}}\simeq m\). 
Namely, at present, the following experimental upper bound is obtained: 
\begin{equation}
m< 0.6 \mbox{ eV}\equiv m\mbox{\tiny max}.\label{astromass2}
\end{equation}
It turns out from Fig.1 that a meaningful bound on the $CP$ violating phase $\beta$ 
\begin{equation}
\sin^2\beta \le \frac{1}{(\sin^2{2\theta\mbox{\tiny solar}})\mbox{\tiny min}}
\left(1-\frac{\langle m_{\nu} \rangle\mbox{\tiny min}^2}{m\mbox{\tiny max}^2}\right)
\end{equation}
is derived for LMA-MSW solution if following condition is satisfied:
\begin{equation}
m\mbox{\tiny max}<\frac{\langle m_{\nu} \rangle\mbox{\tiny min}}{\sqrt{1-(\sin^2{2\theta\mbox{\tiny solar}})\mbox{\tiny min}}}. \label{eq15}
\end{equation}
Thus the constraint on $\beta$ from the present experiments is rather weak. So next we show in Fig.2 how this constraint is restricted as the future experiments make progress on the precision measurement, that is, as the lower bounds of $\langle m_\nu\rangle$ and $\sin^2 2\theta_{\mbox{\tiny solar}}$ are increased.
We also obtain a possible lower bound,  
\begin{equation}
\frac{1}{(\sin^2{2\theta\mbox{\tiny solar}})\mbox{\tiny max}}
\left(1-\frac{\langle m_{\nu} \rangle\mbox{\tiny max}^2}{m\mbox{\tiny min}^2}\right) \le \sin^2\beta, 
\end{equation}
if an experimental lower bound \(m\mbox{\tiny min}\) for \(m\) (i.e. 
\(m\mbox{\tiny min}<m\)) is found in the future experiments and the condition 
\(\langle m_{\nu} \rangle\mbox{\tiny max}<m\mbox{\tiny min}\) is satisfied.
The $\beta$ is not restricted in the case of SMA-MSW solution.
\par
Next, following the method used in Ref\cite{Matsuda}, 
we discuss the bound on \(\sin^2\beta\) by using numerical analysis. 
In the following discussions, we assume \(m_1\alt m_2\alt m_3\). 
The results are scarcely changed for the inverse hierarchical case.
In order to obtain the constraints among the observable quantities,
let us use 
$\overline{m_{\nu}}$, $\Delta m_{12}^2\equiv m_2^2-m_1^2$ and  
$\Delta m_{23}^2\equiv m_3^2-m_2^2$ instead of \(m_1\), \(m_2\) and \(m_3\).
Namely, Inserting the relations, \(m_2=\sqrt{m_1^2+\Delta m_{12}^2}\) 
and \(m_3=\sqrt{m_2^2+\Delta m_{23}^2}=\sqrt{m_1^2+\Delta m_{12}^2+\Delta m_{23}^2}\) into Eq.(\ref{betamass}) with the unitarity condition that 
$|U_{e1}|^2=1-|U_{e2}|^2-|U_{e3}|^2$, 
we obtain the following expressions for $m_1$, $m_2$, and $m_3$ 
\cite{Matsuda}:
\begin{eqnarray}
m_1 & =&
  \sqrt{\overline{m_{\nu}}^2-(|U_{e2}|^2+|U_{e3}|^2)\Delta m_{12}^2-|U_{e3}|^2
  \Delta m_{23}^2}, \nonumber\\
m_2 & =&\sqrt{\overline{m_{\nu}}^2+(1-|U_{e2}|^2-|U_{e3}|^2)\Delta m_{12}^2-|U_{e3}|^2
  \Delta m_{23}^2},\nonumber\\
m_3 & =&
  \sqrt{\overline{m_{\nu}}^2+(1-|U_{e2}|^2-|U_{e3}|^2)\Delta m_{12}^2+(1-|U_{e3}|^2)
  \Delta m_{23}^2}.\label{eq1220-12}
\end{eqnarray}
To show \(\beta\) dependence of  $\overline{m_{\nu}}$, 
we use the center values; 
\(\langle m_{\nu} \rangle=0.39 \mbox{ eV}\), \(|U_{e2}|^2=0.29\) (LMA-MSW solution), 
\(|U_{e3}|^2=0.03\), \(\Delta m_{\mbox{\tiny solar}}^2=4.5\times10^{-5}eV^2 \), and 
\(\Delta m_{\mbox{\tiny atm}}^2=3.2\times10^{-3}eV^2\) as a typical case. By inserting Eq.(\ref{eq1220-12}) into Eq.(\ref{betabetamass}), we get a relation among \(\overline{m_{\nu}}\), \(\beta\), and \(\rho\prime\) as depicted in Fig.3. 
Also from Eq.(\ref{eq1220-12}) and Eq.(\ref{astromass}), we obtain the upper bound for \(\overline{m_{\nu}}\) as
\begin{equation}
\overline{m_{\nu}}<0.6 \mbox{ eV}.\label{astro3} 
\end{equation}
This has been superimposed on Fig. 3 (b) and (c) giving the upper bound of \(\sin^2\beta\) as 
\begin{equation}
\sin^2\beta < 0.7\, .
\end{equation}
Of course, this upper bound depends on the input values of \(\langle m_{\nu} \rangle\), \(|U_{e2}|^2\), \(|U_{e3}|^2\), \(\Delta m_{\mbox{\tiny solar}}^2\), 
and \(\Delta m_{\mbox{\tiny atm}}^2\).
The Fig.3 also predicts the lower limit of the averaged neutrino mass $\overline{m_{\nu}}$ as \(\overline{m_{\nu}}>\langle m_{\nu} \rangle\): Namely, we obtain \begin{equation}
\overline{m_{\nu}}>0.39\mbox{ eV},
\end{equation}
indicating that we have a chance for detecting nonzero 
\(\overline{m_{\nu}}\) in the future beta decay experiment.
\par
In conclusion, 
we have obtained the bounds for the Majorana $CP$ violating phases 
from the recent data of the HEIDELBERG-MOSCOW double beta decay experiment 
incorporating the data from the neutrino oscillation, the astrophysical observation, and the single beta decay experiments.
We have also predicted the lower bound for neutrino mass that may be measurable in the future beta decay experiments.  
\ \\



\begin{figure}[htbp]
\begin{center}
\includegraphics{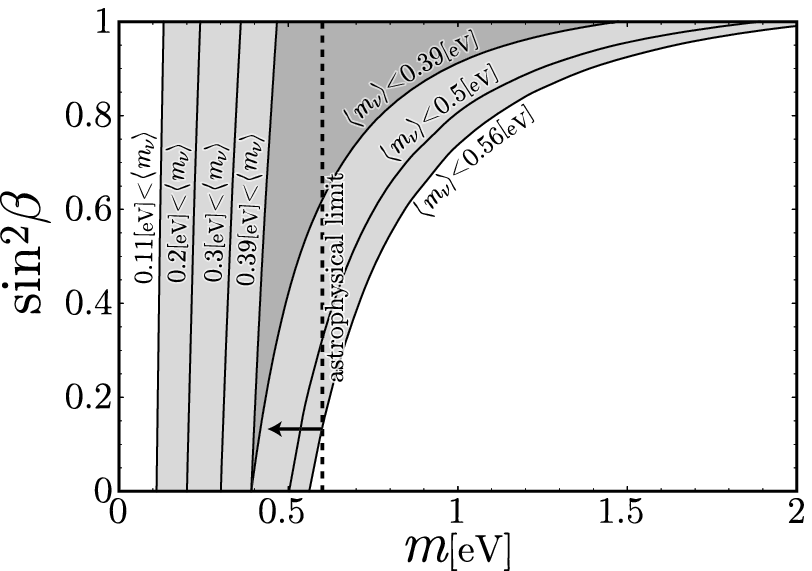}
\end{center}
\caption{%
The allowed regions in the \(\sin^2\beta\)\(-\)\(m\) plane for LMA solutions, Eq.(\ref{eq20501}) with 
\(m \equiv m_1\simeq m_2\). 
The dark shaded region is for the center values and the light ones are for the other empirical values.} 
\label{fig1}
\end{figure}
\begin{figure}[htbp]
\begin{center}
\includegraphics{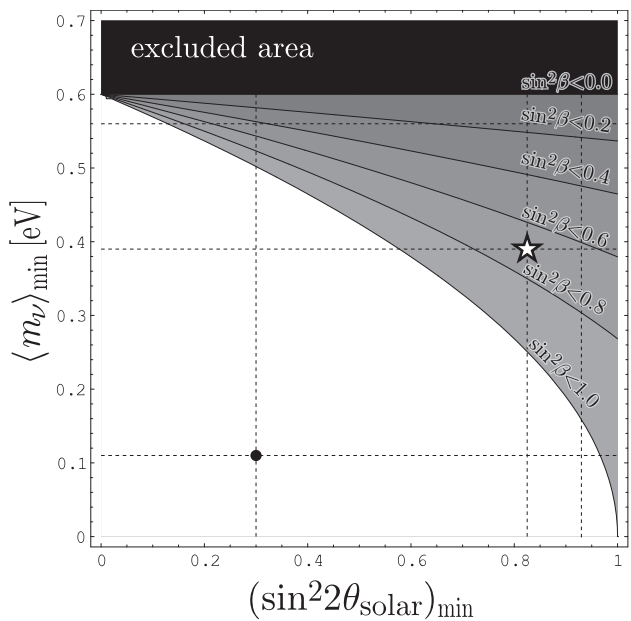}
\end{center}
\caption{The possible upper bounds for \(\sin^2\beta\) shown in the \((\sin^2{2\theta\mbox{\tiny solar}})\mbox{\tiny min}\)\(-\)\(\langle m_{\nu} \rangle\mbox{\tiny min}\) plane. 
The dot indicates the present experimental value, implying that it does not restrict \(\beta\).
The star shows that \(\beta\) is constrained as \(\sin^2\beta<0.7\) if we use the center values.
}
\label{fig2}

\end{figure}
\begin{figure}[htbp]
\begin{center}
\includegraphics{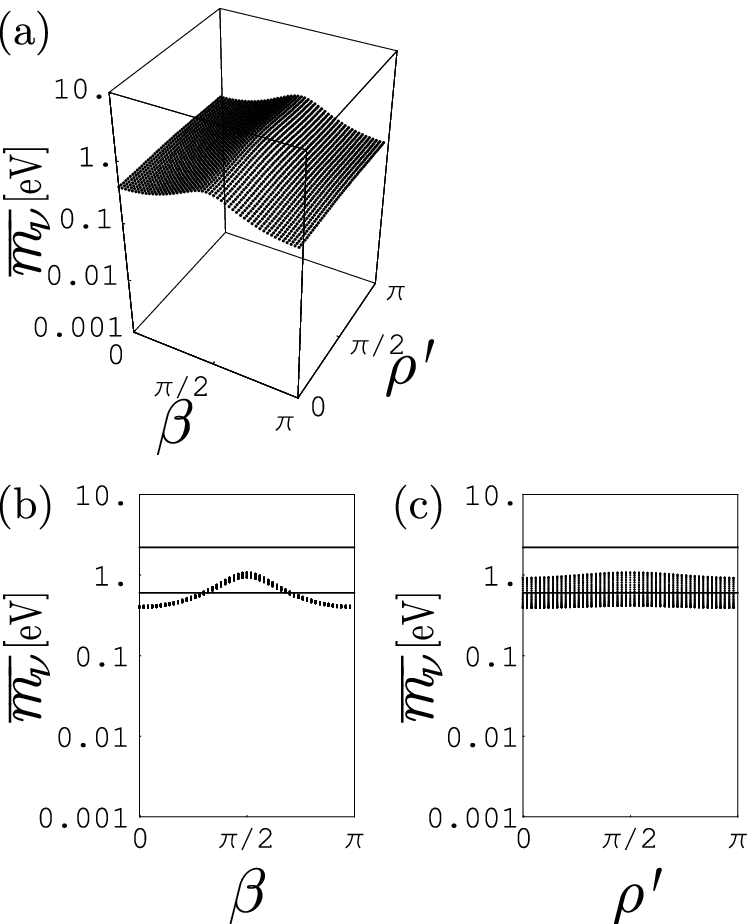}
\end{center}
\caption{The relation among \(\overline{m_{\nu}}\), \(\beta\), and \(\rho\prime\) variables for the LMA-MSW solution for the solar neutrino problem. 
We have fixed the following values: \(\langle m_{\nu} \rangle=0.39\) eV, \(\Delta m_{\mbox{\tiny solar}}^2=4.5\times10^{-5}eV^2 \), \(|U_{e2}|^2=0.29\) (LMA-MSW solution), 
\(\Delta m_{\mbox{\tiny atm}}^2=3.2\times10^{-3}eV^2\), and \(|U_{e3}|^2=0.03\). 
The lower (upper) solid lines in (b) and (c) indicates the upper limit of astrophysics Eq.(\ref{astromass}) (single beta decay Eq.(\ref{betamass})).}
\label{fig3}
\end{figure}


\begin{thebibliography}{99}
\bibitem{skamioka}
T. Kajita, talk presented at Neutrino '98, Takayama, Japan, June 1998; 
H. Sobel, talk presented at Neutrino 2000, Sudbury, Canada, June 2000; 
Y. Fukuda et al, 
Phys. Rev. Lett. {\bf 81}, 1158 (1998); Phys. Rev. Lett. {\bf 81}, 1562 (1998); 
Phys. Lett. {\bf B433}, 9 (1998); Phys. Rev. Lett. {\bf 85}, 3999 (2000); 
Phys. Rev. Lett. {\bf 86}, 5656; 
Q.R. Ahmad, {\it et. al.}, Phys. Rev. {\bf 87}, 071301 (2001).
\bibitem{Matsuda}
K. Matsuda, N. Takeda, T. Fukuyama, and H. Nishiura, Phys. Rev. {\bf D64}, 013001 (2001) 
and references therein.
\bibitem{nishiura}
H. Nishiura, K. Matsuda, T. Kikuchi, and T. Fukuyama, hep-ph/0202189; 
T. Fukuyama and N. Okada, hep-ph/0202214.
\bibitem{Klapdor}
H.V. Klapdor-Kleingrothaus, A. Dietz, H.L. Harney, and I.V. Krivosheina,
Mod. Phys. Lett. {\bf A16}, 2409 (2001).
\bibitem{Klapdor2}
H.V. Klapdor-Kleingrothaus and U. Sarkar, Mod. Phys. Lett. {\bf A16}, 
2469 (2001)(hep-ph/0201224);
V. Barger, S.L. Glashow, D. Marfatia, and K. Whisnant, hep-ph/0201262;
T. Hambye, hep-ph/0201307; H. Minakata and H. Sugiyama, hep-ph/0202003; 
Zhi-zhong Xing, hep-ph/0202034; N. Haba and T. Suzuki, hep-ph/0202143; 
W. Rodejohann, hep-ph/0203214.
\bibitem{doi}
M. Doi, T. Kotani, H. Nishiura, K. Okuda, and E. Takasugi, Phys. Lett. 
{\bf 102B}, 323 (1981).
\bibitem{MNS}
Z. Maki, M. Nakagawa, and S. Sakata, Prog. Theor. Phys. {\bf 28}, 247 (1962).
\bibitem{bilenky}
S.M. Bilenky, J. Hosek and S.T. Petcov, Phys. Lett. {\bf 94B} 495 (1980); 
J. Schechter and J.W.F. Valle, Phys. Rev. {\bf D22} 2227 (1980); 
M. Doi, T. Kotani, H. Nishiura, K. Okuda and E. Takasugi, Phys. Lett. 
{\bf 102B} 323 (1981); A. Barroso and J. Maalampi, 
Phys. Lett. {\bf 132B} 355 (1983).
\bibitem{weinheimer}
Ch. Weinheimer et al., Phys. Lett. {\bf B460} 219 (1999); 
V.M. Lobashev et al., Phys. Lett. {\bf B460} 227 (1999);  
Particle Data Group, Eur. Phys. J. {\bf C15} 1 (2000);
C. Weinheimer, talk presented at Neutrino 2000 
(Sudbury, Canada, June 2000);
V.M. Lobashev, talk presented at Neutrino 2000 
(Sudbury, Canada, June 2000).
\bibitem{fukugita}
M. Fukugita, G.C. Liu, and N. Sugiyama, 
Phys. Rev. Lett. {\bf 84}, 1082 (2000).\\
See also the recent report:
$\Phi$. Elgar$\phi$y et al, arXiv:astro-ph/0204152

\bibitem{chooz}
M. Apollonio et.al., Phys. Lett. {\bf B420}, 397 (1998).


\end{thebibliography}
\end{document}